\documentclass[11pt,fleqn]{article}
\usepackage{amssymb}

\usepackage{amsmath}

\usepackage{amsmath,amssymb,amsthm}

\begin{document}
{\Large
\begin{center}{Reduction of Multidimensional \\ Wave Equations
\\ to Two-Dimensional Equations:\\ Investigation of Possible Reduced
Equations}

\vskip 20pt {\large \textbf{Irina YEHORCHENKO}}

\vskip 20pt {Institute of Mathematics of NAS Ukraine, 3
Tereshchenkivs'ka Str., 01601 Kyiv-4, Ukraine} \\
E-mail: iyegorch@imath.kiev.ua
\end{center}

\vskip 50pt
\begin{abstract} We study possible Lie and non-classical reductions of multidimensional wave equations
and the special classes of possible reduced equations - their symmetries and equivalence classes.
Such investigation allows to find many new conditional and hidden symmetries of the original equations.
\end{abstract}

\section{Why nonlinear wave equation}
We study Lie and non-classical reductions of multidimensional wave equations
and special classes of possible reduced equations - their symmetries and equivalence
classes, as well as types of the reduced equations which represent interesting
classes of two-dimensional equations - parabolic, hyperbolic
and elliptic. This paper carries on the discussion in \cite{yehorchenko:cypr08}.

Ansatzes and methods used for reduction of the d'Alembert ($n$-dimensional wave)
equation can be also used for arbitrary Poincar\'e-invariant equations. Later we will show that
this seemingly simple and partial problem involves many important aspects in the studies
of the PDE.

The topic we consider demonstrates relations of the symmetry methods
(see e.g. \cite{yehorchenko:OvsR}, \cite{yehorchenko:Olver1}) to other aspects
of investigation of PDE - compatibility of systems of equations, methods of finding
general solutions (e.g. by means of hodograph transformations).

The methods we used were not fully algorithmic - it was necessary
to decide when to switch methods, and lot of hypotheses had to be tested.

We consider the multidimensional wave
equation
\begin{gather}\label{yehorchenko:nl wave}
\Box u=F(u),\\
\Box\equiv \partial^2_{x_0}-\partial^2_{x_1}-\cdots
-\partial^2_{x_n}, \quad u=u(x_0, x_1, \ldots, x_n)\nonumber
\end{gather}

It seems thoroughly studied and almost trivial. Let us list only some papers where solutions
of this equation are studied specifically - \cite{yehorchenko:FSS}-\cite{yehorchenko:FBar96}.

However, this equation appears to have many new facets and ideas to discover. Let us point
out that investigation of hyperbolic equations, both with respect to their conditional symmetry
and classification, is considerably more difficult than the same problem for equations in which
at least for one variable partial derivatives have only lower order than the order of equation.

\section{Reduction of nonlinear wave equations - ansatz}
We found conditions of reduction of the multidimensional wave
equation
\[
\Box u=F(u),
\]
by means of the ansatz with two new independent variables.

\begin{equation} \label{yehorchenko:ansatz2}
u=\varphi(y,z),
\end{equation} where $y$, $z$ are new variables.
Henceforth $n$ is the number of independent space variables in the
initial d'Alembert equation.

Reduction conditions for such ansatz are a~system of the d'Alembert and three Hamilton-type
equations:
\begin{gather} \label{yehorchenko:comp conditions}
y_\mu y_{\mu}=r(y,z), \quad y_{\mu} z_{\mu}=q(y,z), \quad
z_{\mu}z_{\mu}=s(y,z), \\
\Box y=R(y, z), \quad \Box z=S(y, z) \nonumber.
\end{gather}

We proved necessary conditions for compatibility of such system of
the reduction conditions . However, the resulting conditions and reduced equations
needed further research.

\section{General background}

Symmetry reduction to equations with smaller
number of independent variables or to ordinary differential
equations (for the algorithms see e.g. the books by Ovsyannikov \cite{yehorchenko:OvsR}
or Olver \cite{yehorchenko:Olver1}).

``Direct method'' (giving wider classes of
solutions than the symmetry reduction) was proposed by P.~Clarkson
and M.~Kruskal \cite{yehorchenko:Clarkson}). See more detailed investigation
of the direct reduction and conditional symmetry in \cite{yehorchenko:FSS},
\cite{yehorchenko:Clarkson}-\cite{yehorchenko:Cicogna-conf03}.
This method for majority of equations results in
considerable difficulties as it requires investigation of
compatibility and solution of cumbersome reduction conditions of
the initial equation.

These reduction conditions are much more
difficult for investigation and solution in the case of equations
containing second and/or higher derivatives for all independent
variables, and for multidimensional equations - e.g. in the
situation of the nonlinear wave equations.

We would like to point out once more that the problem we consider has two specific
difficulties. First, it is always more technically difficult to work with hyperbolic
equations such as the nonlinear wave equation than with parabolic ones (such as evolution equations).
Second, normally the methods and algorithms for working with reductions and exact solutions
are designed and applied for a limited number of variables - usually two or three. Here we work with arbitrary
number of variables, though we limit the number of variables for specific examples.

\section{Compatibility of the reduction conditions: summary}
A similar problem was considered previously for an ansatz
with one independent variable
\begin{equation} \label{yehorchenko:ansatz1}
u=\varphi(y),
\end{equation} where $y$ is a new independent variable.

Compatibility analysis of the d'Alembert--Hamilton system
\begin{gather} \label{yehorchenko:dA-H1}
\Box u=F(u), \quad u_{\mu} u_{\mu}=f(u)
\end{gather}
in the three-dimensional space was done by Collins \cite{yehorchenko:Collins1}.

The sufficient conditions of reduction of the wave equation to an ODE
and the general solution of the system (\ref{yehorchenko:dA-H1}) in the case of three spatial dimensions
 were found by Fushchych, Zhdanov, Revenko \cite{yehorchenko:FZhR-UMZh}.
For discussion of previous results in this area see \cite{yehorchenko:CieciuraGrundland}.
It is evident that the d'Alembert--Hamilton
system~(\ref{yehorchenko:dA-H1}) may be reduced by local
transformations to the form
\begin{gather} \label{yehorchenko:dA-H1-lambda}
\Box u=F(u), \quad u_{\mu} u_{\mu}=\lambda, \quad \lambda=0,\pm1.
\end{gather}
{\bf Statement}\cite{yehorchenko:FZhY}. {\it For the
system~\eqref{yehorchenko:dA-H1-lambda} $(u=u(x_0,x_1,x_2,x_3))$
to be compatible it is necessary and sufficient that the function
$F$ have the following form:}
\[
F=\frac{\lambda}{N(u+C)}, \quad N=0,1,2,3.
\]

Ansatzes of the type (\ref{yehorchenko:ansatz2}) for some particular cases was studied in
\cite{yehorchenko:BarYur}-\cite{yehorchenko:AncoLiu}.

\section{Transformations of compatibility conditions}
Substitution of ansatz $u=\varphi(y,z)$ into the
equation $\Box u=F(u)$ leads to the following
equation (see \cite{yehorchenko:cypr08}):
\begin{gather} \label{yehorchenko:reduction}
\varphi_{yy}y_{\mu}y_{\mu}+2\varphi_{yz}z_{\mu}y_{\mu}+
\varphi_{zz}z_{\mu}z_{\mu}+\varphi_y \Box y+\varphi_z \Box
z=F(\varphi)\\
\left(y_{\mu}=\frac{\partial y}{\partial x_{\mu}}, \ \
\varphi_y=\frac{\partial \varphi}{\partial y}\right),\nonumber
\end{gather}
whence we get a system of equations:
\begin{gather} \label{yehorchenko:comp conditions}
y_\mu y_{\mu}=r(y,z), \quad y_{\mu} z_{\mu}=q(y,z), \quad
z_{\mu}z_{\mu}=s(y,z), \\
\Box y=R(y, z), \quad \Box z=S(y, z) \nonumber.
\end{gather}

System (\ref{yehorchenko:comp conditions}) is a reduction
condition for the multidimensional wave
equation~(\ref{yehorchenko:nl wave}) to the two-dimensional
equation~(\ref{yehorchenko:reduction}) by means of
ansatz $u=\varphi(y,z)$.

The system of equations (\ref{yehorchenko:comp conditions}),
depending on the sign of the expression~$rs-q^2$, may be reduced
by local transformations to one of the following types:

1) elliptic case: $rs-q^2>0$, $v=v(y,z)$~is a complex--valued
function,
\begin{gather}
\Box v=V(v, v^*), \quad \Box v^*=V^*(v, v^*), \nonumber \\
v^*_{\mu}v_{\mu}=h(v, v^*), \quad v_{\mu}v_{\mu}=0,
 \quad v^*_{\mu} v^*_{\mu}=0 \label{yehorchenko:ellipt}
\end{gather}
(the reduced equation is of the elliptic type);

2) hyperbolic case: $rs-q^2 < 0$, $v=v(y, z)$, $w=w(y, z)$~are
real functions,
\begin{gather}
\Box v=V(v, w), \quad \Box w = W(v, w), \nonumber\\
v_{\mu}w_{\mu}=h(v, w), \quad v_{\mu}v_{\mu}=0, \quad w_{\mu}
w_{\mu}=0 \label{yehorchenko:hyperb}
\end{gather}
(the reduced equation is of the hyperbolic type);

3)  parabolic case: $rs-q^2=0$, $r^2+s^2+q^2\not=0$, $v(y,z)$,
$w(y,z)$~are real functions,
\begin{gather}
\Box v=V(v,w), \quad \Box w = W(v,w), \nonumber \\
v_{\mu}w_{\mu}=0, \quad v_{\mu}v_{\mu}=\lambda
 \ (\lambda=\pm 1), \quad w_{\mu} w_{\mu}=0 \label{yehorchenko:parab}
\end{gather}
(if $W\not=0$, then the reduced equation is of the parabolic
type);

4) first-order equations: ($r=s=q=0$), $y \to v$, $z \to w$
\begin{gather}
 v_{\mu}v_{\mu}=w_{\mu} w_{\mu}=v_{\mu}w_{\mu}=0, \nonumber\\
\Box v=V(v, w), \quad \Box w=W(v, w). \label{yehorchenko:1order}
\end{gather}

{\it Elliptic case}

{\bf Theorem 1.} {\it System \eqref{yehorchenko:ellipt} is
compatible if and only if}
\[
V=\frac{h(v,v^*)\partial_{v^*}\Phi}{\Phi}, \quad
\partial_{v^*}\equiv \frac{\partial}{\partial v^*},
\]
{\it where $\Phi$ is an arbitrary function for which the following
condition is satisfied}
\[
(h\partial_{v^*})^{n+1}\Phi=0.
\]

The function $h$ may be represented in the form
$h=\frac{1}{R_{vv^*}}$, where $R$ is an arbitrary sufficiently
smooth function, $R_v$, $R_{v^*}$ are partial derivatives by the
respective variables.

Then the function $\Phi$ may be represented in the form
$\Phi=\sum\limits_{k=0}^{n+1}f_k(v)R_v^k$, where~$f_k(v)$ are
arbitrary functions, and
\[
V=\frac{\sum\limits_{k=1}^{n+1}kf_k(v)R_v^k}{\sum\limits_{k=0}^{n+1}f_k(v)R_v^k}.
\]

The respective reduced equation will have the form
\begin{equation} \label{yehorchenko:ellipt-red}
h(v,v^*)\left(2\phi_{vv^*}+\phi_{v}\frac{\partial_{v^*}\Phi}{\Phi}+\phi_{v^*}\frac{\partial_v\Phi^*}{\Phi^*}\right)=
F(\phi).
\end{equation}

The equation (\ref{yehorchenko:ellipt-red}) may also be rewritten
as an equation with two real independent variables
($v=\omega+\theta$, $v^*=\omega-\theta$):
\begin{equation} \nonumber
\widetilde{h}(\omega,\theta)(\phi_{\omega \omega}+ \phi_{\theta
\theta}) + \Omega(\omega,\theta) \phi_{\omega} +
\Theta(\omega,\theta) \phi_{\theta}= F(\phi).
\end{equation}

{\it Hyperbolic case}

{\bf Theorem 2.} 
{\it System \eqref{yehorchenko:hyperb} is
compatible if and only if}
\[
V=\frac{h(v,w)\partial_{w}\Phi}{\Phi}, \quad
W=\frac{h(v,w)\partial_v\Psi}{\Psi},
\]
{\it where the functions $\Phi$, $\Psi$ are arbitrary functions
for which the following conditions are satisfied}
\[
(h\partial_v)^{n+1}\Psi=0, \quad (h\partial_w)^{n+1}\Phi=0.
\]

The function $h$ may be presented in the
form~$h=\frac{1}{R_{vw}}$, where $R$ is an arbitrary sufficiently
smooth function, $R_v$, $R_w$ are partial derivatives by the
respective variables. Then the functions~$\Phi$, $\Psi$ may be
represented in the form
\[
\Phi=\sum_{k=0}^{n+1}f_k(v)R_v^k, \quad
\Psi=\sum_{k=0}^{n+1}g_k(w)R_w^k,
\]
where $f_k(v)$, $g_k(w)$ are arbitrary functions, ³
\[
V=\frac{\sum\limits_{k=1}^{n+1}kf_k(v)R_v^k}{\sum\limits_{k=0}^{n+1}f_k(v)R_v^k},
\quad
W=\frac{\sum\limits_{k=1}^{n+1}kg_k(w)R_w^k}{\sum\limits_{k=0}^{n+1}g_k(w)R_w^k}.
\]

The respective reduced equation will have the form
\begin{equation} \label{yehorchenko:hyperb-red1}
h(v,w)\left(2\phi_{vw}+\phi_{v}\frac{\partial_w\Phi}{\Phi}+\phi_{w}\frac{\partial_v\Psi}{\Psi}\right)=
F(\phi).
\end{equation}
The equation (\ref{yehorchenko:hyperb-red1}) may also be rewritten
as a standard wave equation
($v=\omega+\theta$, $w=\omega-\theta$):
\begin{equation} \nonumber
\widetilde{h}(\omega,\theta)(\phi_{\omega \omega}- \phi_{\theta
\theta}) + \Omega(\omega,\theta) \phi_{\omega} +
\Theta(\omega,\theta) \phi_{\theta}= F(\phi).
\end{equation}

{\it Parabolic case}

{\bf Theorem 3.} 
{\it System~\eqref{yehorchenko:parab} is
compatible if and only if}
\[
V=\frac{\lambda \partial_v \Phi}{\Phi}, \quad
\partial_v^{n+1}\Phi=0, \quad W\equiv 0.
\]

Equation $\Box u=F(u)$ by means of
ansatz $u=\varphi(y,z)$ cannot be reduced to a
parabolic equation -- in this case one of the variables will enter
the reduced ordinary differential equation of the first order as a
parameter.

System (\ref{yehorchenko:1order}) is compatible only in the case
if $V=W\equiv 0$, that is the reduced equation may be only an
algebraic equation~$F(u)$=0. Thus we cannot reduce
equation $\Box u=F(u)$ by means of
ansatz $u=\varphi(y,z)$ to a first-order equation.

Proof of the above theorems is done by means of the well-known Hamilton--Cayley
theorem, in accordance to which a matrix is a root of its
characteristic polynomial.

\section{Reduction and conditional symmetry}

Solutions obtained by the direct reduction are related to symmetry
properties of the equation -- $Q$-conditional symmetry of this
equation
(symmetries of such type are also called non-classical or non-Lie
symmetries.
It is also possible to see from the previous papers that symmetry of the
two-dimensional reduced equations is often wider than symmetry of
the initial equation, that is the reduction to two-dimensional
equations allows to find new non-Lie solutions and hidden
symmetries of the initial equation (see e.g. papers by Abraham-Schrauner and Leach
\cite{yehorchenko:Abraham}, \cite{yehorchenko:Leach})
The Hamilton equation may also be considered, irrespective of the
reduction problem, as an additional condition for the d'Alembert
equation that allows extending the symmetry of this equation.

Let us look at the wave equation in two spatial dimensions.
Reduction of $\Box u=F(u)$ by our ansatz $u=\varphi(v,w)$
means $Q$-conditional invariance this equation under the operator
\[
Q=\partial_{x_o}+\tau_1(x_o,x_1,x_2)\partial_{x_1}+\tau_2(x_o,x_1,x_2)\partial_{x_2}.
\]
This equivalence of reduction and $Q$-conditional symmetry was proved by Zhdanov, Tsyfra
and Popovych \cite{yehorchenko:zhdanov&tsyfra&popovych99}.
New variables $v$, $w$ are invariants of the operator $Q$:
\[
Qv=Qw=0.
\]

\section{Study of the reduced equations}

Equivalence of quasilinear wave equations is studied well, but we consider a particular class
of such equations.

We consider the reduced equation of the form
\[
h(v,w)\left(2\phi_{vw}+\phi_{v}\frac{\partial_w\Phi}{\Phi}+\phi_{w}\frac{\partial_v\Psi}{\Psi}\right)=
F(\phi).
\]

where the functions $\Phi$, $\Psi$ are arbitrary functions
for which the following conditions are satisfied
\[
(h\partial_v)^{n+1}\Psi=0, \quad (h\partial_w)^{n+1}\Phi=0.
\]

Equivalence transformation of the reduced equations are only of the type
\[
h(v,w)\rightarrow k(v)l(w)h(v,w), v\leftrightarrow w; \phi\rightarrow a\phi+b.
\]
There will be special additional equivalence groups only for special forms of the function $F$.
Special class of the reduced equations - $h(v,w)=k(v)l(w)$; in this case
the equations can be reduced to the case $h(v,w)=const$. All symmetry reductions
have $h(v,w)=const$ and linear $\Phi$, $\Psi$.

We have quite narrow equivalence group of the reduced equation as we actually
took a single representative of an equivalence class of hyperbolic reduced equations.

Description of all possible reductions involves classification of the reductions
found and nomination of certain inequivalent representatives.
Any classification problem is a description of equivalence classes under certain equivalence relations.

Selection of an equivalence group for classification may be in principle arbitrary,
but as a rule one of the following is selected: either the symmetry group of the conditions
describing the initial limited class or the group of automorphisms of some general class.

There is a generally accepted method for classification of symmetry reductions - by subalgebras
inequivalent up to conjugacy. This method does not work for general reductions,
and we have to choose another method of classification.

Another important note - if we do classification in several steps, we have to consider commutativity and
associativity of classification conditions (e.g. under some equivalence group)
adopted at each step.

\section{Example: Solutions for the two-dimensional case}

We will look for parametric solutions for the system
\begin{gather}
\Box v=V(v, w), \quad \Box w = W(v, w), \nonumber\\
v_{\mu}w_{\mu}=h(v, w), \quad v_{\mu}v_{\mu}=0, \quad w_{\mu}
w_{\mu}=0,  \quad \mu=0,1,2 \nonumber
\end{gather}
First we construct parametric or explicit solutions for the equations
$w_{\mu}w_{\mu}=0), v_{\mu}v_{\mu}=0$, and then use them to find solutions of other equations.

Rank 0

General solution of the equations $v_{\mu}v_{\mu}=0$, $w_{\mu}w_{\mu}=0$
\begin{gather}
v=A_{\mu}x_{\mu}+B, w=C_{\mu}x_{\mu}+D;\nonumber \\
A_{\mu}A_{\mu}=0, C_{\mu}C_{\mu}=0. \nonumber
\end{gather}
$p$, $q$ are parametric functions on $x$,

$A_{\mu}$ (${\mu}=1,2$), $B$, $C_{\mu}$ (${\mu}=1,2$), $D$  are arbitrary constants up to conditions.
In this case $h$ will be constant, $\Box v= \Box w=0$ and we will have solutions that can be obtained by symmetry
reduction.

Rank 1

General solution of the equations $v_{\mu}v_{\mu}=0$, $w_{\mu}w_{\mu}=0$
\begin{gather}
v=A_{\mu}(p)x_{\mu}+B(p), w=C_{\mu}(q)x_{\mu}+D(q);\nonumber \\
A^{\mu}_{p}x_{\mu}+B_{p}=0; C^{\mu}_{q}x_{\mu}+D_{q}=0; \nonumber \\
A_{\mu}A_{\mu}=0, C_{\mu}C_{\mu}=0. \nonumber
\end{gather}
$p$, $q$ are parametric functions on $x$,

$A_{\mu}$ (${\mu}=1,2$), $B$, $C_{\mu}$ (${\mu}=1,2$), $D$  are arbitrary functions up to conditions.

Rank 2

General solution of the equations $v_{\mu}v_{\mu}=0$, $w_{\mu}w_{\mu}=0$
\begin{gather} \nonumber
v=A_{\mu}(p_1,p_2)x_{\mu}+B(p_1,p_2), w=C_{\mu}(q_1,q_2)x_{\mu}+D(q_1,q_2);\nonumber \\
A^{\mu}_{p_k}x_{\mu}+B_{p_k}=0; C^{\mu}_{q_k}x_{\mu}+D_{q_k}=0; \nonumber \\
A_{\mu}A_{\mu}=0, C_{\mu}C_{\mu}=0. \nonumber
\end{gather}
$p$, $q$ are parametric functions on $x$,

$A_{\mu}$ (${\mu}=1,2$), $B$, $C_{\mu}$ (${\mu}=1,2$), $D$  are arbitrary functions up to  conditions.

It is easy to prove that that for $v_{\mu}w_{\mu}=h(v, w)$ solutions of $v_{\mu}v_{\mu}=0$,
$ w_{\mu}w_{\mu}=0$, should have the same rank. Further we can find partial parametric
solutions taking the same parameter functions $p$ for $v$ and $w$. This way we will have
new non-Lie solutions with hidden infinite symmetry. (For definition of hidden symmetry see \cite{yehorchenko:Abraham}).

It is well-known \cite{yehorchenko:Sobolev} that the general solution of the system \ref{yehorchenko:dA-H1}
with $F=f=0$, $n=1,2$ can be written as
\begin{gather} \nonumber
u=A_{\mu}(p_1,p_2)x_{\mu}+B(p_1,p_2),\nonumber \\
A^{\mu}_{p_k}x_{\mu}+B_{p_k}=0,\nonumber \\
A_{\mu}A_{\mu}=0, A^{\mu}_{p_k}A^{\mu}_{p_m}=0.\nonumber
\end{gather}

Similarly we can construct a parametric solution for (\ref{yehorchenko:hyperb})
with $V=W=0,h=const$.
\begin{gather} \nonumber
v=A_{\mu}(p_1,p_2)x_{\mu}+B(p_1,p_2),\nonumber \\
A^{\mu}_{p_k}x_{\mu}+B_{p_k}=0,\\
A_{\mu}A_{\mu}=0, A^{\mu}_{p_k}A^{\mu}_{p_m}=0;\nonumber \\
w=C_{\mu}(p_1,p_2)x_{\mu}+D(p_1,p_2)\nonumber \\
C^{\mu}_{p_k}x_{\mu}+D_{p_k}=0,\nonumber \\
C_{\mu}C_{\mu}=0, C^{\mu}_{p_k}C^{\mu}_{p_m}=0;\nonumber \\
A_{\mu}C_{\mu}=const \nonumber.
\end{gather}

The operator of $Q$-conditional symmetry that gives such ansatz will have the form
\begin{gather} \nonumber
Q=\partial_0+\tau_1\partial_1+\tau_2\partial_2,\nonumber \\
\tau_1=\frac{C_0A_2-A_0C_2}{A_1C_2-A_2C_1},
\tau_2=\frac{C_0A_1-A_0C_1}{A_1C_2-A_2C_1}. \nonumber
\end{gather}

\section{Conclusion}
The topic we discuss is closely related to majority of main ideas in the symmetry
analysis of PDE - direct reduction of PDE; conditional symmetry; $Q$-conditional symmetry;
finding solutions directly using nonlocal transformations; group classification of equations and systems of equations;

Our general problem - study of reductions of the nonlinear wave equation (and of other equations
in general) requires several classifications up to equivalence on the way.

At each step we have to define correctly the criteria of equivalence, and check commutativity and associativity
of these equivalence conditions - or otherwise take into account lack of such properties.

\section{Further research}

\begin{enumerate} \vspace{-1mm}\itemsep=-1pt

\item Study of Lie and conditional symmetry of the system of the
reduction conditions.

\item Investigation of Lie and conditional symmetry of the reduced
equations. Finding exact solutions of the reduced equations.

\item Finding of places of previously found exact solutions on the
general equivalence map.

\item Relation of the equivalence group of the class of the
reduced equations with symmetry of the initial equation.

\item Finding and investigation of compatibility conditions and
classes of the reduced equations for other types of equations, in
particular, for Poincar\'e--invariant scalar equations.

\end{enumerate}

\subsection*{Acknowledgements}

I would like to thank the organisers of the Workshop on Group
Analysis of Differential Equations and Integrable Systems for the
wonderful conference and the University of Cyprus and the
Cyprus Research Promotion Foundation (project number
$\Pi$PO$\Sigma$E$\Lambda$KY$\Sigma$H/$\Pi$PONE/0308/01) for support of my participation.


\begin{thebibliography}{99}
\footnotesize

\bibitem{yehorchenko:cypr08} Yehorchenko I.A., Reduction of non-linear
d'Alembert equations to two-dimensional equations,
in Proceedings of the 4th Workshop
"Group Analysis of Differential Equations and Integrable Systems", 2009, p. 243-253

\bibitem{yehorchenko:OvsR}
Ovsyannikov L.V., Group analysis of differential equations,  New
York, Academic Press, 1982.

\bibitem{yehorchenko:Olver1} Olver P., Application of Lie groups to
differential equations.~-- New York: Sprin\-ger Verlag, 1987.

\bibitem{yehorchenko:FSS}
Fushchych~W.I., Shtelen~W.M., Serov~N.I. Symmetry analysis and
exact solutions of nonlinear equations of mathematical physics.~--
Dordrecht: Kluwer Publishers, 1993.

\bibitem{yehorchenko:BlumanKumeiBook} Bluman G.W., Kumei~S.
Symmetries and differential equations.~-- New York: Springer
Verlag, 1989.

\bibitem{yehorchenko:GrunlandHarnadWinternitz94}
Grundland A., Harnad J., Winternitz P., Symmetry reduction for
nonlinear relativistically invariant equations, {\it
J.~Math.~Phys.}, 1984, {\bf 25}, 791--807.

\bibitem{yehorchenko:Tajiri84} Tajiri M., Some remarks on
similarity and soliton solutions of nonlinear Klein-Gordon
equations, {\it J.~Phys.~Soc.~Japan}, 1984, {\bf 53}, 3759--3764.

\bibitem{yehorchenko:PShW} Patera J., Sharp R.T., Winternitz P.
and Zassenhaus H., Subgroups of the Poincar\'e group and their
invariants, {\it  J.~Math.~Phys.}, 1976, {\bf 17}, 977--985.

\bibitem{yehorchenko:FSerovLdAE} Fushchych W.I., Serov N.I.,
The symmetry and some exact solutions of the nonlinear
many-dimensional Liouville, d'Alembert and eikonal equation, {\it
J.~Phys.~A}, 1983, {\bf 16}, 3645--3656.

\bibitem{yehorchenko:FBar} Fushchych~W.~I., Barannyk~A.~F.,
On exact solutions of the nonlinear d'Alembert equation in
Minkowski space $R(1,n)$, {\it Reports of Acad. Sci. of Ukr. SSR,
Ser.A}, 1990, No.~6, 31--34.;

Fushchych W.I, Barannyk~L.~F., Barannyk~A.~F., Subgroup analysis
of Galilei and Poincar\'e groups and reduction of non-linear
equations, Kyiv, Naukova~Dumka, 1991, 304~p. (in Russian);

Fushchych W.I, Barannyk~A.~F., Maximal subalgebras of the rank
$n>1$ of the algebra $AP(1,n)$ and reduction of non-linear wave
equations. I, {\it Ukrain.~Math.~J.}, 1990, {\bf 42}, No.~11,
1250--1256;

Fushchych W.I, Barannyk~A.~F., Maximal subalgebras of the rank
$n>1$ of the algebra $AP(1,n)$ and reduction of non-linear wave
equations. II, {\it Ukrain.~Math.~J.}, 1990, {\bf 42}, No.~12,
1693--1700.

\bibitem{yehorchenko:F95} Fushchych W.I. Ansatz--95,
{\it J.~of~Nonlin.~Math.~Phys.}, 1995, {\bf 2}, 216--235

\bibitem{yehorchenko:FBar96}  Fushchych W.I, Barannyk~A.F.,
On a new method of construction of solutions for non-linear wave
equations, {\it Reports of the Nat. Acad. Sci. of Ukraine}, 1996,
No.~10, 48--51.

\bibitem{yehorchenko:Clarkson} Clarkson, P.A., Kruskal M., New
similarity reductions of the Boussinesq equations, {\it J. Math.
Phys.}, 1989, {\bf 30}, 2201--2213.

\bibitem{yehorchenko:Clarkson Mansfield CS NLWE} Clarkson,~Peter~A.;
 Mansfield,~Elizabeth~L.,
Algorithms for the nonclassical method of symmetry reductions {\it
SIAM~Journal~on Appl.~Math.}, {\bf 54}, No.~6, 1693--1719. 1994,
solv-int/9401002

\bibitem{yehorchenko:Olver94} Olver~P.,  Direct reduction and
differential constraints, {\it Proc.~Roy.~Soc.~London}, 1994, {\bf
A444}, 509--523.

\bibitem{yehorchenko:OlverRosenau} Olver P. and Rosenau P.,
The construction of special solutions to partial differential
equations, {\it Phys.~Lett.~A}, 1986, {\bf 114}, 107--112.

\bibitem{yehorchenko:LeviWinternitz} Levi D. and Winternitz P.,
Non-classical symmetry reduction: example of the Boussinesq
equation, {\it J.~Phys.~A}, 1989, {\bf 22}, 2915--2924.

\bibitem{yehorchenko:zhdanov&tsyfra&popovych99}
Zhdanov~R.Z., Tsyfra~I.M. and Popovych~R.O., A precise definition
of reduction of partial differential equations, {\it
J.~Math.~Anal.~Appl.}, 1999, {\bf 238}, No.~1, 101--123.

\bibitem{yehorchenko:Cicogna-conf03} Cicogna G., A discussion
on the different notions of symmetry of differential equations, in
Proceedinds of Fifth International Conference "Symmetry in
Nonlinear Mathematical Physics" (June 23-29, 2003, Kyiv), Editors
A.G.~Nikitin, V.M.~Boyko, R.O.~Popovych and I.A.~Yehorchenko, {\it
Proceedings of Institute of Mathematics}, Kyiv, 2004, {\bf 50},
Part 1, 77--84.

\bibitem{yehorchenko:Collins1} Collins S.B., Complex potential
equations. I, {\it Math.~Proc.~Camb.~Phil.~Soc.}, 1976, {\bf 80},
165--187;

Collins~S.B., All solutions to a nonlinear system of complex
potential equations, {\it J. Math. Phys.}, 1980, {\bf 21},
240--248.

\bibitem{yehorchenko:FZhR-UMZh} Fushchych W.I, Zhdanov R.Z., Revenko I.V.,
General solutions of nonlinear wave equation and eikonal equation,
{\it Ukrain.~Math.~J.}, 1991, {\bf 43}, No.~11, 1471--1486.

\bibitem{yehorchenko:CieciuraGrundland} Cieciura G., Grundland A.,
A certain class of solutions of the nonlinear wave equation, {\it
J.~Math.~Phys.}, 1984, {\bf 25}, 3460--3469.

\bibitem{yehorchenko:FZhY} Fushchych W.I., Zhdanov R.Z., Yegorchenko I.A., On
reduction of the nonlinear many-dimensional wave equations and
compatibility of the d'Alembert--Hamilton system, {\it
J.~Math.~Anal.~Appl.}, 1991, {\bf 160}, 352--360.

\bibitem{yehorchenko:BarYur} Barannyk A., Yuryk I. On Some Exact Solutions of Nonlinear Wave
Equations, in Proceedings of the Second International Conference
"Symmetry in Nonlinear Mathematical Physics" (7-13~July~1997,
Kyiv), 1997, Editors M.I.~Shkil', A.G.~Nikitin and V.M.~Boyko,
Institute~of~Mathematics, Kyiv, {\bf 1}, 1997, 98--107.
\bibitem{yehorchenko:zhdanov-panchakBoxu}
Zhdanov R. and Panchak Olena, New conditional symmetries and exact
solutions of the nonlinear wave equation, {\it J.~Phys.~A}, 1998,
{\bf 31}, 8727-8734
\bibitem{yehorchenko:Euler CS NLWE} Euler~Marianna and
Euler~Norbert, Symmetries for a class of explicitly space- and
time-dependent (1+1)-dimensional wave equations. {\it
J.~Nonlin.~Math.~Phys.} 1997, {\bf 1}, 70--78.
\bibitem{yehorchenko:AncoLiu} Anco~S.C., Liu S.,
Exact solutions of semilinear radial wave equations in $n$
dimensions, {\it J.~Math.~Anal.~Appl.}, 2004, {\bf 297}, 317--342,
math-ph/0309049.

\bibitem{yehorchenko:Abraham}Abraham-Shrauner B., Hidden
symmetries and nonlocal group generators for ordinary differential
equations, {\it IMA J.Appl.Math.}, 1996, V.56, 235--252;
Abraham-Shrauner B., Hidden symmetries, first integrals and
reduction of order of nonlinear ordinary differential equations,
{\it J. Nonlin. Math. Phys.}, 2002, V.9, Suppl.~2, 1--9.

\bibitem{yehorchenko:Leach}Abraham-Shrauner B. and Leach P.G.L., Hidden Symmetries of Nonlinear Ordinary
Differential Equations, in Exploiting Symmetry in Applied and Numerical Analysis,
Lectures in Applied Mathematics.,  American Mathematical Society, 1993, V. 29,
1–10.

\bibitem{yehorchenko:Sobolev} Sobolev S.L.: Functional-invariant solutions to the wave equation (Russian).
{\it Tr. Steklov Phys. Mat. Inst.}, 1934, V.5, 259-264;

\end{thebibliography}
\end{document}